# 3D Finite Element Analysis of HMA Overlay Mix Design to Control Reflective Cracking

Ziad Georges Ghauch*

*Undergraduate student, Department of Civil Engineering, Lebanese American University,*

*Blat, Byblos – Lebanon,*

*Corresponding author, E-mail: zdghaouche@gmail.com

This study examines the effectiveness of HMA overlay design strategies for the purpose of controlling the development of reflective cracking. A parametric study was conducted using a 3D Finite Element (FE) model of a rigid pavement section including Linear Viscoelastic (LVE) material properties for the Hot Mix Asphalt (HMA) overlay and non-uniform tire-pavement contact stresses. Several asphalt mixtures were tested in the surface, intermediate, and leveling course of the HMA overlay. Results obtained show that no benefits can be anticipated by using either Polymer-Modified (PM) or Dense-Graded (DG) mixtures instead of Standard Binder (SB) mixtures in the surface or intermediate course. For the leveling course, the use of a PM asphalt binder was found beneficial in terms of mitigating reflective cracking. As compared to the SB mix, the use of PM asphalt mixture in the leveling course reduced the level of longitudinal tensile stress at the bottom of the HMA overlay above the PCC joint by approximately 30%.

**Keywords:** Reflective cracking, PCC pavements, HMA overlay, Finite Element analysis, Viscoelastic constitutive model.

**Introduction**

*Background*

When an old rigid pavement reaches the end of its service life, a common rehabilitation measure is to place an HMA overlay on the existing pavement. However, after a relatively short period of time, cracks appear at the surface of the new HMA overlay in a pattern similar to that of the existing cracks and/or joints of the old pavement. This phenomenon is known as reflective cracking, or reflection cracking. Reflective cracking is a fatigue process mainly due to overlays being subjected to millions of traffic wheel loads as well as daily and seasonal temperature cycling. The primary mechanisms behind reflective cracking are horizontal and vertical movements in the underlying old pavement at the vicinity of joints/cracks (Bozkurt 2002). Vehicle wheel loads approaching the PCC crack create both vertical differential displacements, or from a fracture mechanics point of view, Mode II cracking (in-plane shearing), and horizontal movements, or Mode I (opening) cracking. In addition, daily and seasonal temperature drops may cause Mode I opening of the PCC crack edges due to thermal shrinkage. Finally, Mode III (tearing mode) cracking sometimes occurs when the vehicle travels alongside a longitudinal crack (Monismith 1980).

The occurrence of reflective cracking implies a reduction in the service life of the HMA overlay, but does not necessarily mean the pavement structure has failed. Among other things, reflective cracking causes a loss of surface water-tightness and higher deformations throughout the pavement structure. In addition, reflective cracking may eventually cause a loss of serviceability and ride quality associated with striping of asphalt concrete along surface cracks.

In order to address the problem of reflective cracking, several empirical, mechanistic, and mechanistic-empirical (M-E) design methods have been proposed. While empirical methods use only field data to associate the pavement performance to design parameters, mechanistic methods use entirely mechanistic procedures to do so. M-E design methods emerge in between, combining past field data with mechanistic analyses. Eltahan and Lytton (2000) presented a M-E design approach against reflective cracking, based on an integration of Paris' law for estimating the number of Equivalent Single Axle Loads (ESAL) for crack propagation. Another M-E design method was proposed by Sousa et al. (2002). The computed Von-Mises strain at the crack tip was associated with the number of ESALs to crack propagation by using Strategic Highway Research Program (SHRP) beam bending fatigue test.

Several treatments have been proposed in order to control the development of reflective cracking. Examples of pre-overlay treatments include increasing overlay thickness (Sherman 1982), cracking and seating (Rajagopal 2004) use of open-graded HMA mixtures (Hensley 1980). The use of interlayer systems has also proven successful in mitigating reflective cracking. Interlayer systems include steel reinforcement to compensate the lack of HMA tensile strength (Elseifi and Al-Qadi 2005), stress-relief interlayer systems such as Interlayer Stress Absorbing Composite – ISAC (Bozkurt and Buttlar 2002) and Stress Absorbing Membrane Interlayer – SAMI to absorb excessive deformations in the cracking zone, and fracture tolerant interlayer systems such as sand –mix (Baek 2010), used for the purpose of increasing the fracture resistance of HMA materials. The effectiveness of any treatment in mitigating reflective cracking depends primarily on the condition of the existing pavement structure. Button and Lytton (2007) observed that current treatment strategies only delay the occurrence and do not prevent the development of reflective cracking.

*Objective and Scope*

The purpose of this study is to examine the effectiveness of several HMA overlay mix design strategies for the purpose of mitigating the development of reflective cracking. In order to achieve this purpose, the HMA overlay was divided into three courses: surface, intermediate, and leveling course, each with a thickness of 50 mm. Then, three asphalt mixture types, namely Dense Graded-DG, Standard Binder-SB, and Polymer Modified-PM were tested for each course, resulting in a total of 27 pavement simulations. Results of the parametric study can be used as a roadmap in the selection of appropriate HMA mixture types that would best delay the occurrence of reflective cracking.

This study is organized as follows. First, a description of the constructed 3D FE pavement model is presented along with the distribution of vertical contact pressure at the tire-pavement interface. Then, the properties of the materials used are reviewed, and the LVE model for HMA materials is described in detail. Results of the performed parametric study are presented after that. Finally, the results are complemented by an analysis of the effect of each asphalt mixture in each course on the stress state at the bottom of the HMA overlay.

**Finite Element Model**

*3D FE Pavement Model*

A 3D FE model was constructed using the FE software ABAQUS (2011). As shown in Figure 1a and 2, the modeled rigid pavement section consists of a 150 mm HMA overlay placed on top of a 220 mm PCC slab, resting on a 180 mm aggregate subbase and 2880 mm subgrade soil. The dimensions of the modeled pavement section were 3.6 m in the longitudinal direction and 2.4 m in the transverse direction. Due to longitudinal symmetry, only half of the pavement structure was modeled. Roller support boundary conditions were assigned to the longitudinal and lateral vertical planes, while fixed supports were assigned to the bottom horizontal plane. In order to achieve optimum accuracy without increase in computational cost, a biased mesh was implemented. Small elements (10mm) were used in the HMA layer at the vicinity of the PCC joint and along the wheel path where high stress-strain gradients occur, and increasingly larger elements were used far away from the joint and loading path. In total, 54,812 eight-node linear brick elements with reduced integration (C3D8R) were used to model the pavement section.

*Moving Vehicle Load*

An 80 KN Single Axle Load (SAL) with dual-tire assembly was modeled in this study. The footprint of one tire of the dual-tire assembly is 220 mm in the longitudinal traffic direction, and 240 mm in the transverse direction, as shown in Figure 1b.

It is well established that tire-pavement contact stresses include, in addition to vertical stresses, longitudinal and lateral shear stresses. The effect of shear stresses on the pavement response is practically limited to the surface where the shear stresses are applied. Hence, for bottom-up reflective cracking analysis, longitudinal and transverse shear stresses can be safely neglected.

The distribution of vertical contact stresses at the tire-pavement interface was assumed to follow a semi-elliptical pressure distribution, given as follows:

$$p(x) = p_{max}\sqrt{1-\left(\frac{x}{a}\right)^2} \qquad (1)$$

Where x is the longitudinal distance along the tire footprint, $p_{max}$ = 1.2 MPa, and 2a is the longitudinal length of the tire footprint. In order to simulate the movement of the load at a speed of 8 km/h, a quasi-static approach was implemented by gradually shifting the contact area of the load along the loading path. A total of 170 increments, each with a 9 milliseconds duration and a length of 20 mm, were used to model one full wheel load passage.

*Material Properties*

HMA materials exhibit time and temperature dependent properties. At low temperatures and high frequency loading, HMA materials behave like elastic solids, while they behave as viscous fluid at high temperatures and low frequency loading. At intermediate temperatures and loading frequencies, these materials behave viscoelastically, presenting a certain degree of elastic rigidity along with a dissipation of energy through frictional losses. Several researchers have successfully used a linear viscoelastic (LVE) constitutive model to represent HMA properties (Baek 2011, Dave and Buttlar 2010). Elseifi et al. (2006) concluded that the elastic theory, as compared to the LVE theory, underestimates the pavement response, therefore pinpointing the need for a viscoelastic constitutive model. In this study, the HMA overlay is considered as a linear isotropic viscoelastic material, with a time-dependent stress represented as follows (Abaqus 2011):

$$\sigma(t) = \int_0^t 2G(\tau - \tau')\dot{e}\,d\tau' + \int_0^t K(\tau - \tau')\dot{\phi}\,d\tau' \qquad (2)$$

Where $\dot{\phi}$ and $\dot{e}$ are the mechanical volumetric and deviatoric strains, respectively, K and G are the bulk and shear modulus, function of the reduced time τ.

The Generalized Maxwell Model was used to represent the viscoelastic behavior of the HMA overlay. As shown in Figure 3, the model is composed of a finite number [M] of Maxwell units and one spring element connected in parallel. A Maxwell unit is composed of a spring and dashpot connected in series to model elastic and viscous behavior, respectively. Each spring element is assigned a relaxation modulus $E_i$, while each viscous dashpot is assigned a friction resistance $\eta_i$. The relaxation function of the generalized Maxwell model is the following:

$$E(\xi) = E_0 - \sum_{i=1}^{M} E_i \left[ e^{-\xi/\tau_i} \right] \quad (3)$$

Where $E_i$ is the i[th] spring elastic modulus, $\tau_i$ is the relaxation time of the i[th] Maxwell unit defined as $\tau_i = \eta_i/E_i$, and $\xi = t/a_T$ is the reduced time, function of the time t and a time-temperature shift factor $a_T$. This equation is known as the Prony (or Dirichlet) series expansion. Using the Prony series expansion, the shear and bulk relaxation moduli are expressed in function of reduced time as follows:

$$G_R(t) = G_0 - \left[ 1 - \sum_{i=1}^{N} g_i \left( 1 - e^{-\tau/\tau_i} \right) \right] \quad (4)$$

$$K_R(t) = K_0 - \left[ 1 - \sum_{i=1}^{N} k_i \left( 1 - e^{-\tau/\tau_i} \right) \right] \quad (5)$$

Where $k_i$ and $g_i$ are dimensionless Prony series parameters, N is the number of parameters, $K_0$ and $G_0$ are the instantaneous bulk and shear relaxation moduli, respectively, and $K_R(t)$ and $G_R(t)$ are the bulk and shear relaxation moduli, respectively.

Values of dimensionless shear relaxation moduli ($g_i$) and corresponding relaxation time ($\tau_i$) were obtained from Al-Qadi et al. (2009) and are presented in Figure 3. The corresponding properties of each binder course are presented in Table 1.

All other material characteristics required for the model are assumed to exhibit isotropic linear elastic behavior, with typical values of Young's modulus (E) and Poisson's ratio (ν) selected from the literature (see Figure 1). The Poisson's ratio of HMA materials is assumed constant with time (υ = 0.35).

**Parametric Study**

The HMA overlay was divided into three courses: surface, intermediate, and leveling course; three asphalt mixture types (i.e. Dense Graded - DG, Standard Binder - SB, and Polymer Modified - PM) were tested for each course, resulting in a total of 27 pavement simulations.

Results of the parametric study are shown in Table 2. It can be noted that values of longitudinal stress (S11) and shear stress (S12) at the bottom of the HMA overlay above the PCC joint result from one full passage of vehicle load. It was observed that as the asphalt binder type was changed in each course, the level of shear stress was not found to vary as much as that of the longitudinal stress.

In terms of longitudinal tensile stress, the lowest value (437.35 KPa) was recorded when all surface, base, and leveling courses are composed of PM asphalt mixtures. Relatively low tensile stresses were also observed (1) when the leveling and base course were composed of PM asphalt mixtures, while the surface course consisted of DG mixture, and (2) when the leveling and surface course were composed of PM asphalt mixtures, while the base course consisted of DG mixture. Such results show that the use of PM asphalt mixtures in the three courses of the HMA overlay, particularly in the leveling course, is extremely beneficial in reducing the longitudinal tensile stresses at the bottom of the HMA overlay above the PCC joint. For comparative purposes, the level of longitudinal stress drops by 54.27 % with the use of PM mixtures instead of SB mixtures in all three courses of the HMA overlay.

As for the shear stresses, the lowest value (491.88 KPa) was observed in the HMA overlay consisting of SB surface and intermediate courses, and PM leveling course. Switching from PM to DG mixtures in the leveling course, or from SB to DG in the intermediate course slightly increased the shear stress to 496.05 KPa and 499.87 KPa respectively, as shown in Table 2.

**Results**

*Surface Course*

The use of Polymer Modified (PM) or Dense Graded (DG) surface course is beneficial in reducing the tensile stresses at the bottom of the HMA overlay. Figure 4 shows the effect of varying the surface course mixture type on the tensile and shear stress response at the bottom of the HMA overlay above the PCC joint. The bars in the histogram of [6 show the average of the values obtained, while the error bars show the minimum and maximum values recorded. The average level of tensile stress dropped by 9.04 % and 8.70 % as the SB surface mixture was replaced by PM and DG mixtures. On the contrary, the lowest level of shear stress was recorded with the SB surface course. In fact, the use of PM and DG mixtures in the surface course slightly increased the shear stress at the bottom of the overlay, by approximately 2.56 % and 2.38 %, respectively.

When a SB asphalt mixture is used in the surface course, relatively higher levels of tensile stress are obtained, as opposed to DG and PM mixtures, while the opposite was true in terms of shear stress. Hence, no specific conclusion as to the effectiveness of the surface course mixture type in reducing the stresses at the bottom of the HMA overlay can be established.

*Intermediate Course*

Using a DG or PM asphalt mixture in the intermediate (base) course is effective in reducing the level of tensile stress at the bottom of the HMA overlay above the PCC joint. As shown in Figure 5, using PM or DG mixtures instead of SB mixtures in the intermediate course lowers the tensile stress by 10.04 % and 9.27 %, respectively. However, in terms of shear stress, the use of SB asphalt mixture in the intermediate course is slightly more efficient than the corresponding DG and PM mixtures, as shown in Figures 5.

For the intermediate course, as well as the surface course, it was found that the advantages of using a DG or PM asphalt mixture over a SB mixture (in terms of reduced tensile stress) are counterbalanced by the occurrence of relatively higher levels of shear stress. Therefore, it can be concluded that no benefits can be anticipated by using either PM or DG mixtures instead of SB mixture in the intermediate course.

*Leveling Course*

As for the leveling course, the use of a PM asphalt binder was found extremely beneficial in terms of both tensile and shear stresses (see Figure 6). As compared to the DG mixture, the PM mixture reduces the level of both tensile and shear stresses by 8.94% and 0.81%, respectively. When compared to the SB mixture, the PM asphalt mix also reduced tensile stress by 29.65% and shear stress by 4.64%, respectively.

Therefore, the use of a PM asphalt mixture in the leveling course is beneficial in reducing both tensile and shear stresses at the bottom of the HMA overlay, and hence reduce the potential for reflective cracking development.

**Conclusions**

This study examines the possibility of modifying HMA overlay mix design in order to mitigate the development of reflective cracking. A parametric study was conducted using a 3D Finite Element (FE) model of a rigid pavement section developed using the FE code ABAQUS v6.11. The pavement model includes a Linear Viscoelastic (LVE) model for Hot Mix Asphalt (HMA) materials and non-uniform tire-pavement contact stresses. The HMA overlay was divided into three courses, and three asphalt mixture types were investigated for each course.

Results obtained suggest that the use of PM asphalt mixture in the HMA overlay, in particular in the leveling course, reduces the longitudinal tensile stresses at the bottom of the overlay. However, use of PM mixtures in the surface or intermediate course slightly increases the shear stresses. Hence, no direct quantification of the effectiveness of the surface or intermediate course mixture type on reducing the potential for reflective cracking development can be established. Using PM mixtures in the leveling course significantly reduces both tensile and shear stresses at the bottom of the HMA overlay. Hence, placing a PM mixture in the leveling course lowers both longitudinal and shear stresses above the PCC joint and therefore reduces the potential for reflective cracking development.

The benefits of using polymer-modified HMA mixtures in terms of improved pavement durability, better performance under extreme temperature fluctuations, and improved resistance against rutting are well known. This study further showed that the use of polymer-modified HMA binders, particularly in the leveling course, reduces the potential for bottom-up reflective cracking. This promotes the idea that even though polymer-modified HMA binders cost relatively more than conventional binders, the advantages of using such mixtures compensate for the additional costs.

**Acknowledgements**

I would like to present special recognition to two persons, Dr. Mazen Tabbara and Dr. Gebran Karam from the department of Civil Engineering at the Lebanese American University for their support and thoughtful guidance. Also, special thanks to Professor Imad Al-Qadi from the University of Illinois at Urbana-Champaign for providing the valuable data for characterizing HMA materials.

Table 1 – Properties of HMA mixtures tested by Al-Qadi et al. (2009)

|    | Binder       | Asphalt Content (%) |
|----|--------------|---------------------|
| DG | SBS PG-22    | 5.4                 |
| PM | SBS PG 70-22 | 4.5                 |
| SB | PG 64-22     | 4.5                 |

Table 2 - Tabulated results of tensile stress S11 (KPa) and shear stress S12 (KPa) obtained from parametric study

| | | | Surface Course | | | | | | | | |
|---|---|---|---|---|---|---|---|---|---|---|---|
| | | | DG | | | PM | | | SB | | |
| | | | Leveling Binder Course | | | | | | | | |
| | | | DG | PM | SB | DG | PM | SB | DG | PM | SB |
| Base Course | DG | S11 | 484.04 | 442.70 | 579.34 | 481.74 | 440.05 | 577.42 | 530.40 | 488.56 | 628.70 |
| | | S12 | 515.69 | 511.49 | 535.84 | 516.56 | 512.61 | 536.61 | 504.17 | 499.87 | 523.29 |
| | PM | S11 | 480.01 | 438.97 | 574.49 | 478.73 | 437.35 | 573.62 | 527.08 | 485.54 | 624.52 |
| | | S12 | 518.41 | 514.29 | 538.52 | 519.14 | 515.27 | 539.15 | 507.02 | 502.79 | 526.11 |
| | SB | S11 | 533.16 | 489.96 | 632.92 | 531.91 | 488.36 | 632.02 | 572.41 | 528.85 | 674.71 |
| | | S12 | 508.22 | 504.15 | 526.48 | 509.22 | 505.40 | 527.38 | 496.05 | 491.88 | 513.43 |

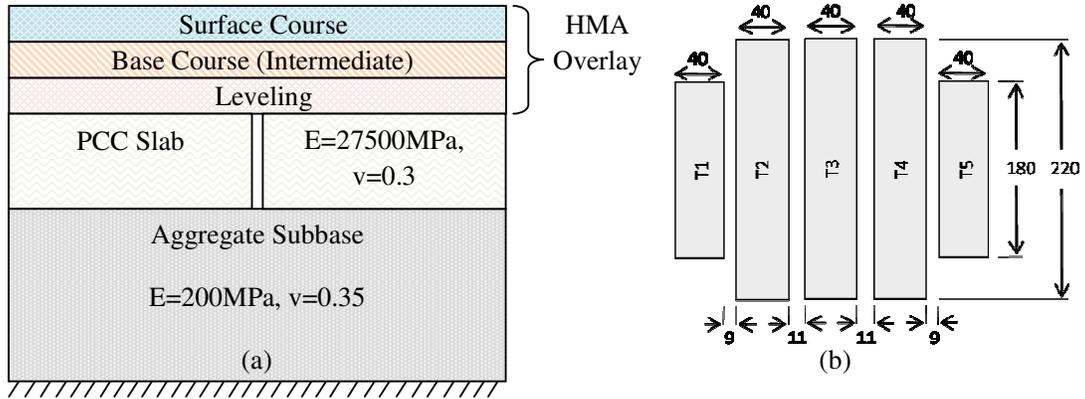

Figure 1 – (a) Profile of modeled pavement section, and (b) modeled tire footprint of one tire in a dual-tire assembly (dimensions in mm)

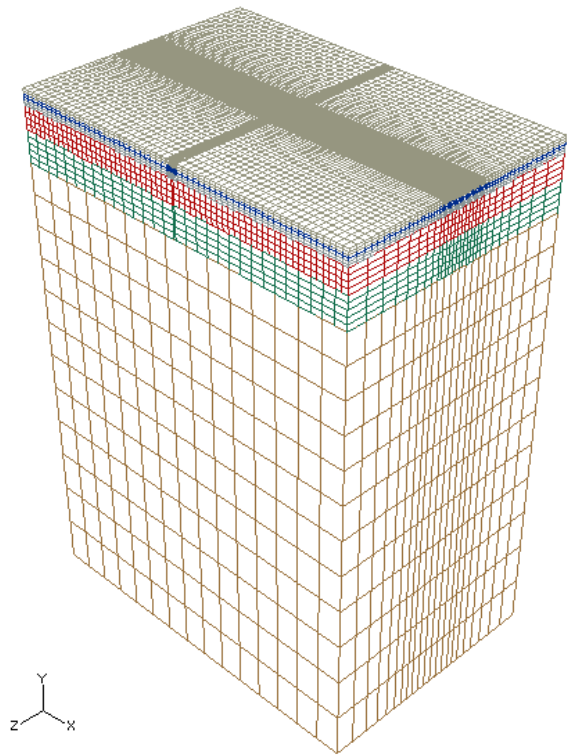

Figure 2 – 3D Finite Element mesh of modeled pavement section

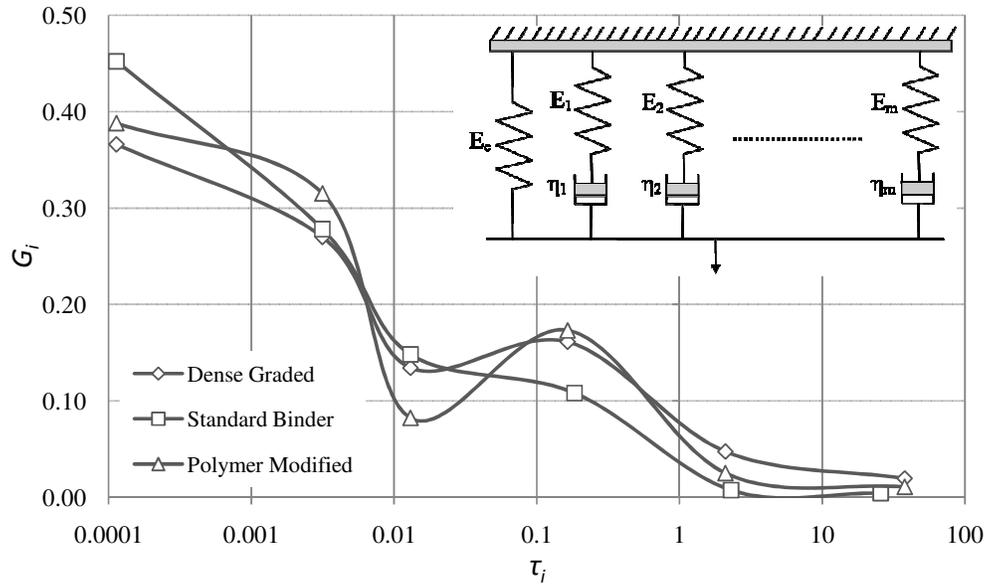

Figure 3 - Prony Series parameters [after Al-Qadi et al. (2009)] and schematic illustration of Generalized Maxwell model

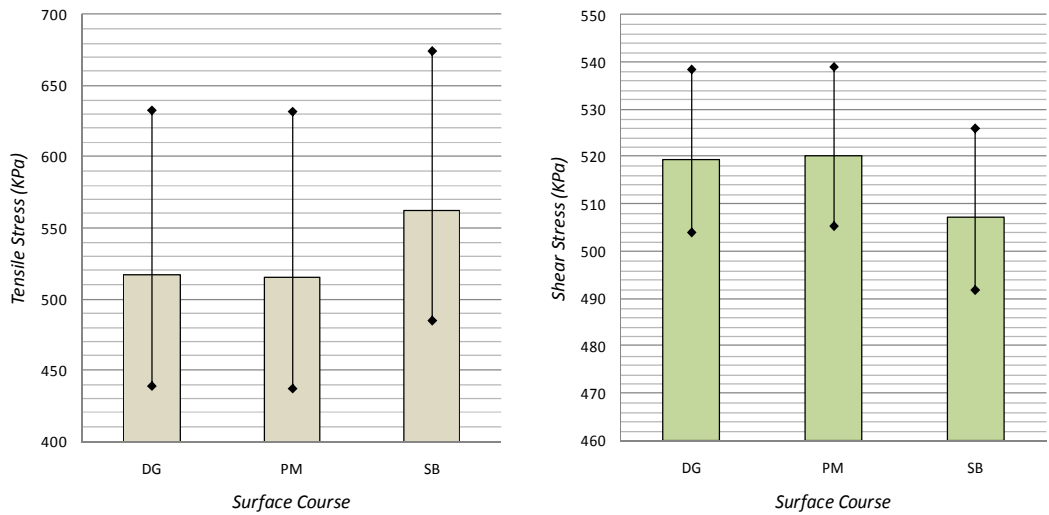

Figure 4 - Effect of surface course asphalt mixture type on (a) tensile stress, and (b) shear stress at the bottom of the HMA overlay

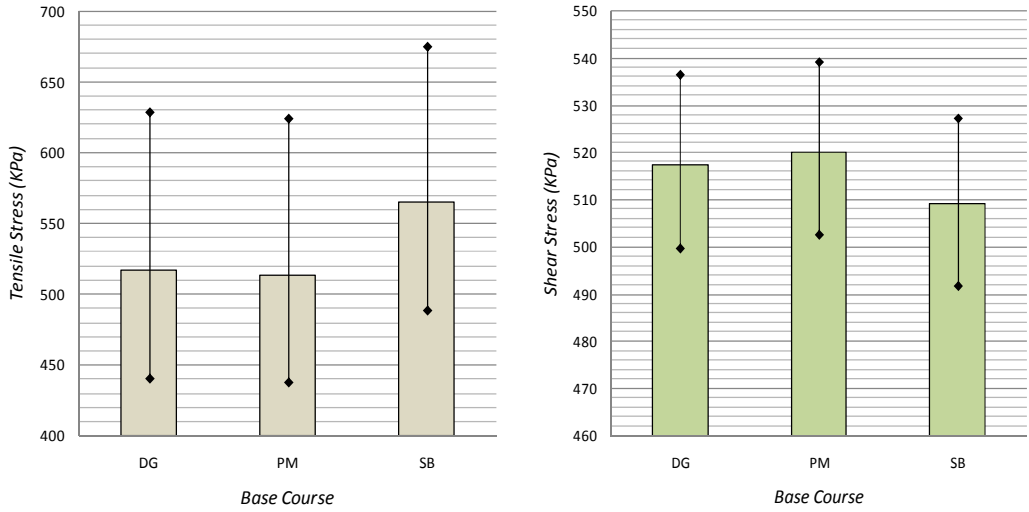

Figure 5 - Effect of base course asphalt mixture type on (a) tensile stress, and (b) shear stress at the bottom of the HMA overlay

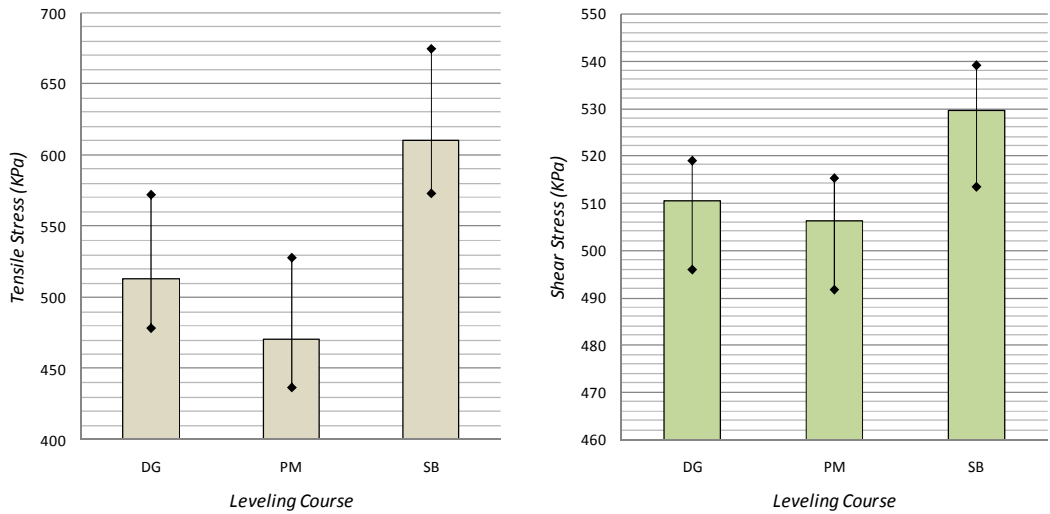

Figure 6 - Effect of leveling course asphalt mixture type on (a) tensile stress, and (b) shear stress at the bottom of the HMA overlay